\documentclass[runningheads]{llncs}

\usepackage[T1]{fontenc}
\usepackage[utf8]{inputenc}
\usepackage[nolist]{acronym}
\usepackage{url}
\def\UrlOrds{\do\*\do\-\do\~\do\'\do\"\do\-}\makeatletter
\g@addto@macro{\UrlBreaks}{\UrlOrds}
\makeatother
\usepackage{hyperref}
\usepackage{booktabs}
\usepackage{multirow}
\usepackage{color,soul}
\usepackage{balance}
\usepackage{tabularx}
\usepackage[table,xcdraw]{xcolor}
\usepackage{xcolor}
\usepackage{blindtext}
\usepackage{mdframed}
\usepackage{subcaption}
\usepackage{tikz}

\usepackage[misc]{ifsym}

\usepackage{graphicx}
\usepackage[export]{adjustbox}

\newcommand\copyrighttext{%
	\footnotesize The final publication is available at Springer via \url{https://doi.org/10.1007/978-3-030-98785-5_24}}
	\newcommand\copyrightnotice{%
	\begin{tikzpicture}[remember picture,overlay]
	\node[anchor=south,yshift=10pt] at (current page.south) {\fbox{\parbox{\dimexpr\textwidth-\fboxsep-\fboxrule\relax}{\copyrighttext}}};
	\end{tikzpicture}%
}

\newcommand{\eg}{e.g., }

\newcommand{\ie}{i.e., }

\begin{acronym}
    \acro{DNS}{Domain Name System}
    \acro{EDNS}{Extended \ac{DNS}}
    \acro{DoUDP}{DNS over UDP}
    \acro{DoTCP}{DNS over TCP}
    \acro{RA}{RIPE Atlas}
    \acro{DoT}{DNS over TLS}
    \acro{DoH}{DNS over HTTPS}
    \acro{DoQ}{DNS over QUIC}
    \acro{DoDTLS}{DNS over DTLS}
    \acro{GSO}{Generic Segmentation Offload}
    \acro{GRO}{Generic Receive Offload}
    \acro{DoS}{Denial of Service}
    \acro{QoE}{Quality of Experience}
    \acro{EU}{Europe}
    \acro{NA}{North America}
    \acro{AF}{Africa}
    \acro{SA}{South America}
    \acro{OC}{Oceania}
\acro{PoP}{Point of Presence}
    \acro{UDP}{User Datagram Protocol}
    \acro{TCP}{Transmission Control Protocol}
    \acro{TLS}{Transport Layer Security}
    \acro{APNIC}{Asia Pacific Network Information Centre}
    \acro{RT}{Response Time}
    \acro{TFO}{TCP Fast Open}
    \acro{FQDN}{Fully Qualified Domain Name}
    \acro{RTT}{Round-Trip Time}
    \acro{ISP}{Internet Service Provider}
    \acro{RCODE}{Response Code}
    \acro{TTL}{Time to Live}
    \acro{CPE}{Customer-premises equipment}
    \acro{IAD}{Integrated access device}
    \acro{WG}{Working Group}
    \acro{IETF}{Internet Engineering Task Force}
    \acro{dnsop}{Domain Name System Operations}
    \acro{RR}{Resource Record}
    \acro{IP}{Internet Protocol}
    \acro{MTU}{Maximum Transmission Unit}
    \acro{RD}{Recursion Desired}
    \acro{ALPN}{Application-Layer Protocol Negotiation}
    \acro{IANA}{Internet Assigned Numbers Authority}
    \acro{SNI}{Server Name Indication}
\end{acronym}

\acrodefplural{AS}[ASes]{Autonomous Systems}
\acrodefplural{TLD}[TLDs]{Top-Level Domains}
\acrodefplural{TTFB}[TTFBs]{Times to First Byte}
\acrodefplural{PoP}[PoPs]{Points of Presence}
\acrodefplural{FQDN}[FQDNs]{Fully Qualified Domain Names}
\acrodefplural{RCODE}[RCODEs]{Response Codes}
\acrodefplural{RR}[RRs]{Resource Records}
\acrodefplural{RT}[RTs]{Response Times}

\begin{document}

\title{One to Rule them All?\newline A First Look at DNS over QUIC}
\titlerunning{One to Rule them All? A First Look at DNS over QUIC}

\authorrunning{M. Kosek et al.}
\tocauthor{Mike Kosek, Trinh Viet Doan, Malte Granderath, Vaibhav Bajpai}

\author{Mike Kosek\inst{1}\textsuperscript{(\Letter)} \and Trinh Viet Doan\inst{1} \and Malte Granderath\inst{1} \and Vaibhav Bajpai\inst{1,2}}
\institute{Technical University of Munich, Germany\\ \and CISPA Helmholtz Center for Information Security, Germany
\email{\{kosek,doan,grandera\}@in.tum.de}, \email{bajpai@cispa.de}}

\maketitle

\vspace{-1em}

\begin{abstract}
    The DNS is one of the most crucial parts of the Internet.
    Since the original DNS specifications defined UDP and TCP as the underlying transport protocols, DNS queries are inherently unencrypted, making them vulnerable to eavesdropping and on-path manipulations.
    Consequently, concerns about DNS privacy have gained attention in recent years, which resulted in the introduction of the encrypted protocols DNS over TLS (DoT) and DNS over HTTPS (DoH).
    Although these protocols address the key issues of adding privacy to the DNS, they are inherently restrained by their underlying transport protocols, which are at strife with, \eg IP fragmentation or multi-RTT handshakes --- challenges which are addressed by QUIC.
    As such, the recent addition of DNS over QUIC (DoQ) promises to improve upon the established DNS protocols.
    However, no studies focusing on DoQ, its adoption, or its response times exist to this date --- a gap we close with our study.
    Our active measurements show a slowly but steadily increasing adoption of DoQ and reveal a high week-over-week fluctuation, which reflects the ongoing development process: As DoQ is still in standardization, implementations and services undergo rapid changes.
    Analyzing the response times of DoQ, we find that roughly 40\% of measurements show considerably higher handshake times than expected, which traces back to the enforcement of the traffic amplification limit despite successful validation of the client's address.
    However, DoQ already outperforms DoT as well as DoH, which makes it the best choice for encrypted DNS to date.
    \copyrightnotice

\end{abstract}

\section{Introduction}
\label{sec:introduction}

The \ac{DNS} is used for almost all communications across the Internet.
As the original DNS specifications~\cite{rfc1034,rfc1035} define UDP and TCP as the underlying transport protocols, DNS requests and responses using \ac{DoUDP} and \ac{DoTCP} are inherently unencrypted, which makes them vulnerable to eavesdropping and on-path manipulations~\cite{rfc9076}.
This enables an observer to not only reveal the browsing or application usage behavior~\cite{connection.oriented.dns}, but also the identification of device types which are in use~\cite{hardaker2018analyzing}; hence, a user profile can be created and tracked with only having access to the user's DNS traffic~\cite{KimZ15,KirchlerHLK16,LiMGLZLG18}.
Consequently, concerns on DNS privacy have gained attention in recent years.

With the standardization of \ac{DoT}~\cite{rfc7858} in 2016 and \ac{DoH}~\cite{rfc8484} in 2018, encrypted DNS protocols leveraging \ac{TLS} on top of TCP have been introduced.
Moreover, \ac{DoDTLS}~\cite{rfc8094} has also been standardized as an experimental protocol in 2017, offering encrypted DNS by leveraging \ac{TLS} on top of UDP.

However, while these protocols address the key issues of adding privacy to DNS~\cite{lu2019dns,deccio2019dns,doan2021measuring,garcia2021large}, they are inherently restrained by their underlying transport protocols.
Using UDP as a connectionless protocol, \ac{DoDTLS} is vulnerable to \emph{IP fragmentation}~\cite{truncation,dns.fragmentation,ietf-dnsop-avoid-fragmentation,domain.validation} --- a problem which has gained awareness in recent years due to the trend of increasing DNS response sizes~\cite{nlnet.dnssec,rssac.data.icann}.
Although both \ac{DoT} and \ac{DoH} are not affected by \emph{IP fragmentation}, as they leverage the connection-oriented TCP, the underlying TCP connections are still constrained by head-of-line-blocking and missing multiplexing support on the transport layer, as well as an additional connection establishment in comparison to UDP.

These challenges are addressed by QUIC~\cite{rfc9000,rfc9001,rfc9002}, a connection-oriented encrypted transport protocol using UDP as a substrate.
Standardized in early 2021, QUIC features mandatory encryption, solves head-of-line blocking, provides multiplexing, and improves on connection establishment time by combining the transport and encryption handshakes into a single round trip.
Consequently, offering DNS using the QUIC transport protocol is the natural evolution for not only the traditional performance-oriented DNS protocols \ac{DoUDP} and \ac{DoTCP}, but also the privacy-preserving DNS protocols \ac{DoT} and \ac{DoH} (as well as the experimental \ac{DoDTLS}).

\ac{DoQ} is currently being standardized within the \emph{DNS PRIVate Exchange} IETF working group~\cite{dprive} with the design goal to provide DNS privacy with minimum latency.
With this objective, DoQ aims to obsolete all other currently used DNS protocols, which lack privacy and/or require more round-trips for handshakes --- therefore, promising to make DoQ the \emph{``One to Rule them All''}.
Despite its development status, multiple experimental implementations already exist that offer \ac{DoQ} support for clients~\cite{dnslookup,quicdog,aioquic,nextdnscliclient,routedns}, servers~\cite{adguardcoredns,adguardhome,quicdog,aioquic}, proxies~\cite{adguarddnsproxy,nextdnscliclient,routedns}, as well as multipurpose libraries~\cite{adguarddnslibs,quicdog,aioquic,flamethrower}.
Moreover, \emph{AdGuard}~\cite{adguard} and \emph{nextDNS}~\cite{nextdns} already use \ac{DoQ} in production systems for their DNS-based ad as well as tracker blocking services, offering publicly reachable \ac{DoQ} servers and client implementations~\cite{dnslookup,nextdnscliclient}.
However, while \ac{DoQ} was submitted to the Internet Engineering Steering Group (IESG) for publication in December 2021, only one study~\cite{garcia2021large} has explicitly included DoQ as part of an experiment on encrypted DNS based on traffic flow analyses as of January 2022.
Hence, no studies focusing on \ac{DoQ}, its adoption, or its response times exist to this date --- a gap we close with our study.

We begin by investigating the adoption of \ac{DoQ} (see \autoref{sec:adoption}) and identify a maximum of 1,217 resolvers in a single week. Over the course of 29 weeks, we find 1,851 unique X.509 certificates used by the resolvers.
However, only 51.6\% of the resolvers in the first week are still reachable in the last week, reflecting the ongoing development and standardization process, during which DoQ implementations and services undergo rapid changes.
Analyzing the response times of DoQ in comparison to \ac{DoUDP}, \ac{DoTCP}, \ac{DoT}, as well as \ac{DoH} (see \autoref{sec:performance}), we find that QUIC's full potential is only utilized in around 20\% of measurements.
On the other hand, roughly 40\% of measurements show considerably higher handshake times than expected, which traces back to the enforcement of the traffic amplification limit despite successful validation of the client's address, ultimately causing an additional, unnecessary round-trip.

The remainder of this paper is structured as follows:
We first present our methodology in \autoref{sec:methodology}.
Afterwards, we detail our adoption measurements in \autoref{sec:adoption} before analyzing the response times of DoQ in \autoref{sec:performance}.
Limitations and future work are discussed in \autoref{sec:limitations}, after which we conclude the paper with \autoref{sec:conclusion}. \section{Methodology}
\label{sec:methodology}

To study the adoption and response times of \acf{DoQ}, we issue measurements from a single vantage point located in the research network of the Technical University of Munich, Germany.
Distributed measurement platforms such as RIPE Atlas do currently not support DoQ; nevertheless, we plan to distribute our measurements to multiple vantage points in the future (see~\autoref{sec:limitations}).

\textbf{Adoption.}
To assess the adoption of DoQ on resolvers worldwide, we issue weekly scans of the IPv4 address space over the course of 29 weeks, starting in 2021-W27 (July 05--11).
For this, we leverage the \emph{ZMap} network scanner~\cite{the_zmap_project} and target all DoQ ports proposed by the different \ac{DoQ} Internet-Drafts~(\mbox{I-Ds}), \ie \texttt{UDP/784}, \texttt{UDP/853}, and \texttt{UDP/8853}~\cite{ietf-dprive-dnsoquic}. 
For comparison, we additionally target DoUDP port \texttt{UDP/53}, which we identify by leveraging the \emph{ZMap}'s built-in DNS probing packet that queries an \texttt{A} record for \texttt{www.google.com}~\cite{zmap_udp_probes}.
Since \emph{ZMap} does not provide means for the identification of QUIC or DoQ, we issue a custom packet~\cite{doq_zmap} that carries the \texttt{Initial} QUIC handshake frame with an invalid version number of \texttt{0}~\cite{rueth.poese.dietzel.hohlfeld.2018}:
In this way, if the target operates a QUIC stack on the probed port, a \texttt{Version Negotiation} packet is triggered.
As such a packet does not produce state, it allows us to identify the target as \emph{QUIC-capable} without consuming resources on the target itself~\cite{rfc9000}.
However, note that other QUIC services, which are not necessarily DoQ, could be offered on the probed ports.
Hence, we further validate targets identified as \emph{QUIC-capable} by the \emph{ZMap} scans, checking if they actually support DNS over QUIC~\cite{doq_verify}.
To do so, we offer the \texttt{doq} \texttt{\ac{ALPN}} identifiers (as required by the DoQ I-Ds~\cite{ietf-dprive-dnsoquic}), which results in a list of \emph{DoQ-capable} targets.
As a final step, a connection to every \emph{DoQ-capable} target on all proposed DoQ ports \texttt{UDP/784}, \texttt{UDP/853}, as well as \texttt{UDP/8853}, is established~\cite{doq_misc}:
For these connections, we offered the QUIC version \texttt{draft-34} in our \texttt{Initial} frame until 2021-W42, while support for version \texttt{1} was added in 2021-W43.
Overall, our client supports the QUIC versions \texttt{draft}\texttt{-34}, \texttt{-32} and \texttt{-29} since the start of our study, as well as version \texttt{1} later on; hence, the client can respond to \texttt{Version Negotiation} packets if issued by the resolvers.
For DoQ, we offer versions in the order of \texttt{draft-06} to \texttt{draft-00}~\cite{ietf-dprive-dnsoquic}, for which we added support for new versions within 2 weeks of the \texttt{draft} release.
By issuing the highest QUIC and DoQ protocol versions supported by our client first, we ensure that we negotiate the highest shared protocol versions between our client and the target resolver.
With this, we record the negotiated QUIC and DoQ versions, as well as the X.509 certificate offered by each \emph{DoQ-capable} target, creating the final list of \emph{DoQ-verified} resolvers.

\textbf{Response Time.}
To study the response times of DoQ compared to \ac{DoUDP}, \ac{DoTCP}, \ac{DoT}, and \ac{DoH}, we develop \emph{DNSPerf}, an open-source DNS measurement tool which supports all stated protocols~\cite{doq_dnsperf}.
Using \emph{DNSPerf} to target all \emph{DoQ-verified} resolvers identified in 2022-W02, we issue response time measurements every hour over the course of 2022-W03 (January 17--23).
As we specifically scan for DoQ in our adoption measurement, we measure \ac{DoUDP}, \ac{DoTCP}, \ac{DoT}, as well as \ac{DoH} \emph{optimistically}, \ie without prior knowledge whether the target resolvers offer the respective DNS protocols in addition to DoQ.
In detail, we measure DoUDP and DoTCP according to RFC 1034~\cite{rfc1034} on target port \texttt{UDP/53} and, respectively, \texttt{TCP/53}, DoT according to RFC 7858~\cite{rfc7858} preferring TLS version \texttt{1.3} on target port \texttt{TCP/853}, and DoH according to RFC 8484~\cite{rfc8484}, also preferring TLS version \texttt{1.3} on target port \texttt{TCP/443}.
Similar to the verification step of the adoption measurement, we again target all proposed DoQ ports.

Our DNS requests query an \texttt{A} record for \texttt{test.com}.
We further explicitly set the \texttt{Recursion Desired} flag in all requests to ensure that the resolvers return a valid and recursively queried or cached \texttt{A} record, which circumvents resolvers simply returning the corresponding name server or refusing to answer our queries to prevent \emph{Cache Snooping}~\cite{trufflehunter,rd_bit_refuse} when the flag is not set.
As populating the caches can affect the measured response times, every DNS request on every protocol is preceded by an identical cache warming query, which ensures that the actual DNS response time measurement query is directly answered by the resolver from a cached record.
For DoQ, we additionally use the negotiated QUIC \texttt{Version} along with the token received in a \texttt{New\_Token} frame of the cache warming query for the handshake of the subsequent DNS response time measurement query, which ensures that the response time measurements are not affected by QUIC's \texttt{Version Negotiation} and \texttt{Address Validation} features.
Overall, these decisions enable comparable DNS response time measurements of all stated protocols.

\textbf{Round-Trip Time (RTT).}
If the response time measurement of a protocol:resolver pair is successful, we measure the \emph{round-trip time} to the targets to analyze the path and protocol-specific \emph{RTT}~\cite{doq_measurements}.
Since the resolvers can be deployed behind proxies, or the path can have protocol-dependent queuing characteristics, we send probing packets to the same port using the same protocol as the respective response time measurement. 
Therefore, our implementation enables \emph{RTT} measurements for UDP (DoUDP), TCP (DoTCP, DoT, DoH), as well as QUIC (DoQ) by leveraging protocol-specific probing payloads:
For UDP, a randomized payload is sent from a random \texttt{Source Port}, while the payload for TCP is a \texttt{SYN} packet containing a randomized \texttt{Sequence Number}.
Finally, QUIC leverages the custom packet that carries the \texttt{Initial} QUIC handshake with an invalid version number of \texttt{0} (see Adoption above).

\textbf{Ethical Considerations.}
To minimize the impact of our active scans, we follow best practices of the Internet measurement community~\cite{zmap.ethics,irtf-pearg-safe-internet-measurement}.
Thus, we display the intent of our scans on a website reachable via the IP address of each scanning machine, also allowing targets to opt-out from our study.
Moreover, we honor opt-out requests from previous studies and maintain a University-wide shared blocklist with the excluded targets.

\textbf{Reproducibility.}
In order to enable the reproduction of our findings~\cite{reproducibility}, we make the developed tools, the raw data of our measurements, as well as the analysis scripts and supplementary files publicly available~\cite{doq_open_source}. \section{Adoption}
\label{sec:adoption}

To study the adoption of DoQ on resolvers worldwide, we issue weekly scans of the IPv4 address space over the course of 29 weeks, as described in~\autoref{sec:methodology}.
Thus, we record the negotiated QUIC and DoQ versions, as well as the X.509 certificates offered by the target resolvers that support DoQ, for which we also determine the announcing \acp{AS} and geolocations.
Overall, we find 1,851 unique X.509 certificates over the course of 29 weeks.\\

\textbf{Adoption of QUIC and DoQ Versions.}
In our scans and in the verification process, we target all proposed DoQ ports \texttt{UDP/784}, \texttt{UDP/853}, and \texttt{UDP/8853}.
The DoQ drafts \texttt{-00} and \texttt{-01} state that port \texttt{UDP/784} \emph{MAY} be used for experimentation.
\texttt{draft-02} defined \texttt{UDP/8853} for usage as experimentation as well as for reservation at the \ac{IANA}.
This was changed in \texttt{draft-03}, where port \texttt{UDP/784} was again stated for experimentation usage; ultimately, \texttt{UDP/853} has been established as the final port for reservation at \ac{IANA}.
Over the course of the 29 weeks, we observe a dominance of the usage of port \texttt{UDP/784}, with roughly 75--82\% of all \emph{DoQ-verified} resolvers offering all observed DoQ \texttt{drafts}\texttt{-00}, \texttt{-02}, and \texttt{-03} on \texttt{UDP/784}.
Port \texttt{UDP/8853} is only observed in combination with \texttt{draft-02} at roughly 17--24\% of all \emph{DoQ-verified} resolvers, with the remainder (<1\%) serving DoQ \texttt{draft-02} on port \texttt{UDP/853}.

\begin{figure*}[t]
	\vspace{-1em}
	\centering
	\includegraphics[width=\linewidth]{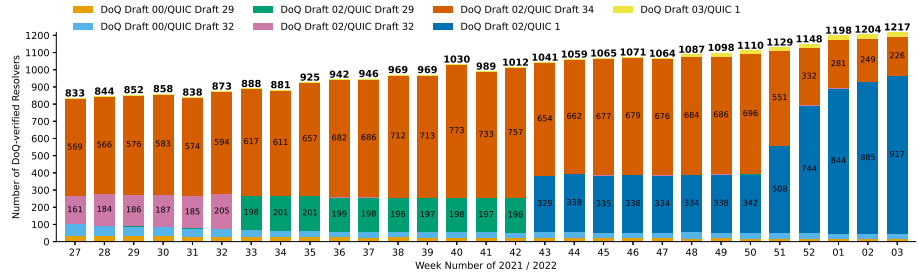}
	\vspace{-1em}
	\caption{Number of \emph{DoQ-verified} resolvers per week number of 2021 and 2022 grouped by negotiated DoQ and QUIC version.
	Support for QUIC version \texttt{1} was added in 2021-W43.
	} \label{fig:adoption}
	\vspace{-1em}
\end{figure*}

Fig.~\ref{fig:adoption} presents the \emph{DoQ-verified} resolvers per week, grouped by negotiated DoQ and QUIC version.
Overall, we observe that the number of \emph{DoQ-verified} resolvers rises steadily: Starting with 833 resolvers in 2021-W27 (July 05--11), we see an increase by 46.1\% to 1,217 verified resolvers in 2022-W03 (January 17--23).
After we added support for QUIC version 1~\cite{rfc9000} in 2021-W43, we observe a steady usage of \texttt{DoQ Draft 02/QUIC 1} (dark blue bars) until 2021-W50, followed by a steep increase until 2022-W01.
Analyzing this observation, we find that the open source DNS server implementation \emph{AdGuard Home (AGH)}~\cite{adguardhome} changed the default DoQ/QUIC pair from \texttt{DoQ Draft 02/QUIC Draft 34} (orange bars) to \texttt{DoQ Draft 02/QUIC 1} (dark blue bars) starting 2021-W51~\cite{adguardhome107}, matching the pattern we observe.
In addition, we find indications of the usage of \emph{AGH} by the updated resolvers within the \emph{Common Names} of their X.509 certificates, and also identify multiple of the updated resolvers to be running \emph{AGH} through random sampling.
Hence, we attribute the observed increase in usage of \texttt{DoQ Draft 02/QUIC 1} between 2021-W51 and 2022-W01 to this implementation.

Although we offer a total combination of 28 DoQ/QUIC version pairs as of 2022-W03 (see~\autoref{sec:methodology}), we observe only 7 pairs across all measurements, with the majority being \texttt{DoQ Draft 02/QUIC 1} (dark blue bars, 917 (75.3\%)) in 2022-W03.
Additionally, we find that only 430 (51.6\%) of the initial 833 resolvers are still verified in 2022-W03.
As a comparison, 96.5\% of the verified DoUDP resolvers from 2021-W27 are still verified in 2022-W03.
This fluctuation of DoQ reflects the development process: While DoQ is still in standardization, implementations and services change frequently and are expected to be used in experimental rather than in production environments.

However, both \emph{AdGuard}~\cite{adguard} and \emph{nextDNS}~\cite{nextdns} actually do use DoQ in production systems for their DNS-based ad and tracker blocking services, offering publicly reachable \ac{DoQ} servers as well as client implementations~\cite{dnslookup,nextdnscliclient}.
This is reflected in the \emph{Common Names} of the X.509 certificates offered by the verified DoQ resolvers:
In 2022-W03, 199 resolvers (16.5\%) state \texttt{dns.nextdns.io} as their common name.
Analyzing the change over time, we observe that \emph{nextDNS} operates the highest share of resolvers in each week, with a mean of roughly 180 resolvers in 2021-W27 to 2021-W31, increasing to a mean of 199 resolvers in 2021-W32 to 2022-W03.
While the increase was observed between 2021-W31 and 2021-W32, \emph{nextDNS} offered \texttt{DoQ Draft 02/QUIC Draft 32} (purple bars) until 2021-W32 and downgraded all resolvers to \texttt{DoQ Draft 02/QUIC Draft 29} (green bars) in 2021-W33, where this DoQ/QUIC pair is exclusively offered by \emph{nextDNS}.
After adding support for QUIC version 1 in 2021-W43, we also observe that all \emph{nextDNS} resolvers offer \texttt{DoQ Draft 02/QUIC 1} (dark blue bars) since that week; hence, we attribute the previously observed downgrade to the missing support of QUIC version 1 in our tooling during that timeframe.
Considering the publicly reachable \ac{DoQ} servers of \emph{AdGuard} (identified by the common names \texttt{dns.adguard.com} and \texttt{adguard.ch}), we identify 25 resolvers offering \texttt{DoQ Draft 03/QUIC 1} (yellow bars) in 2022-W03 (2.1\%).
Note that this DoQ/QUIC pair is exclusively offered by the \emph{AdGuard} services, as it differs from the \emph{AdGuard Home (AGH)} open source DNS server implementation detailed above. 
We find 12--17 resolvers with the common name \texttt{dns.adguard.com} and \texttt{DoQ Draft 02/QUIC Draft 34} (orange bars) until 2021-W47, after which these resolvers switch to \texttt{DoQ Draft 03/QUIC 1} (yellow bars) starting 2021-W48.
Moreover, \texttt{DoQ Draft 03/QUIC 1} is also offered by 6--8 resolvers using \texttt{adguard.ch} starting 2021-W49.\\

\begin{figure*}[t]
	\vspace{-1em}
	\centering
	\includegraphics[width=0.75\linewidth,trim=0.27cm 1.54cm 0.27cm 0.4cm, clip, frame]{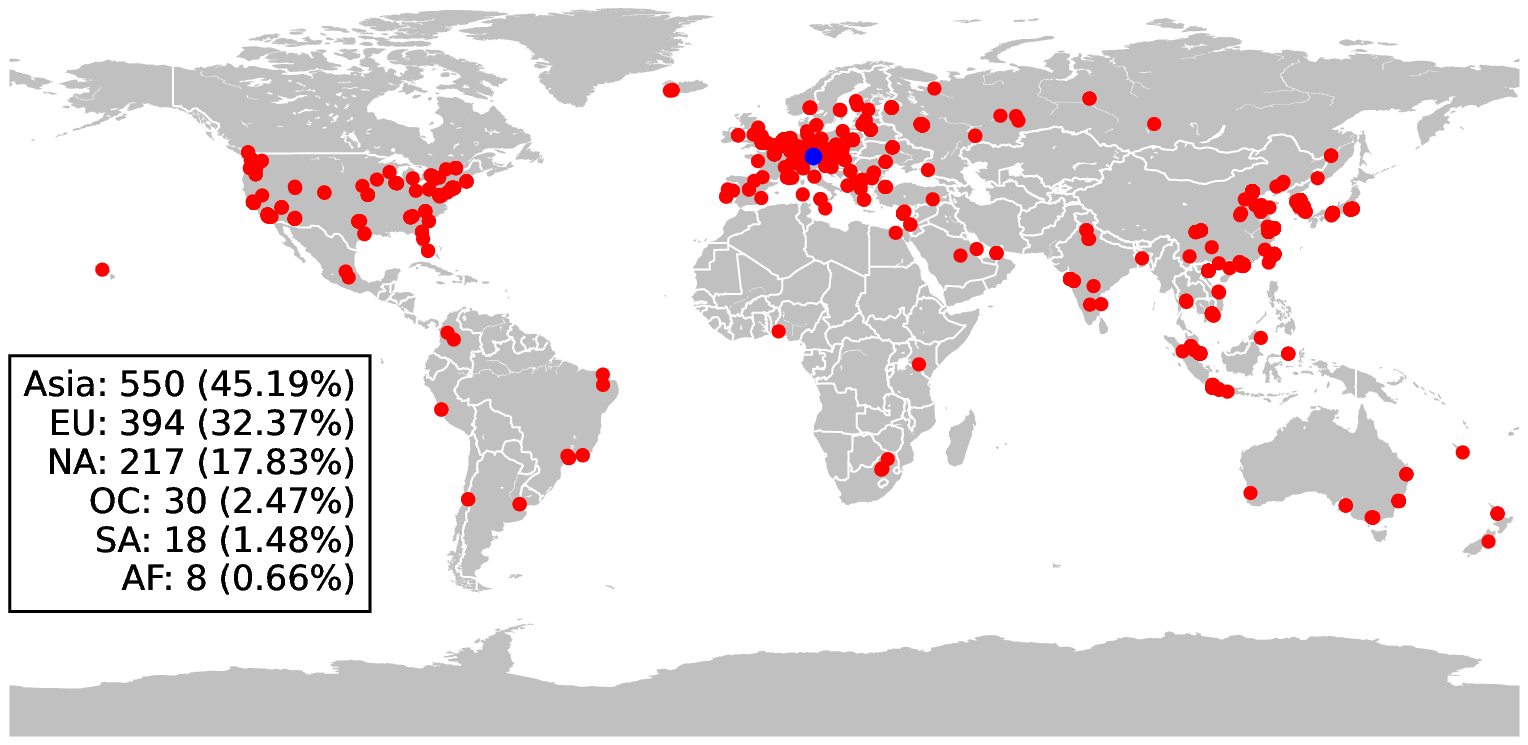}
	\caption{Geographical locations of the 1,217 \emph{DoQ-verified} resolvers as of 2022-W03, with counts by continent. The blue marker represents our vantage point.
	} \label{fig:doq-resolver-map}
	\vspace{-1em}
\end{figure*}
 
\textbf{Adoption in Continents and by ASes.}
Figure \ref{fig:doq-resolver-map} presents the geographical locations of the 1,217 \emph{DoQ-verified} resolvers of 2022-W03 with counts per continent based on an IPv4 geolocation lookup service~\cite{ipapi}.
We observe a strong focus of resolvers operated in Asia (45.19\%) and \ac{EU} (32.37\%), whereas only 17.83\% are operated in \ac{NA}.
However, note that geolocation lookups of IP addresses are known to have inaccuracies, possibly resulting in the incorrect attribution of locations.

The publicly available information of \emph{AdGuard}~\cite{adguardoverview} states that they operate resolvers in 10 countries in the four continents Asia, \ac{EU}, \ac{NA}, as well as \ac{OC}.
However, this is not reflected in our measurements: We find that 16 resolvers are operated in Russia (\ac{EU}, MNGTNET (AS199274)), 8 in Cyprus (Asia, ADGUARD (AS212772)), and 1 in Italy (\ac{EU}, TISCALI-IT (AS8612)) for 2022-W03, resulting in an overall distribution over 2 continents, 3 countries, as well as 3 ASes.
Due to the strong divergence, we attribute this observation to the incorrect attribution of the IP geolocation lookups.

On the other hand, \emph{nextDNS} operates globally distributed DoQ resolvers:
The 199 \emph{DoQ-verified} resolvers of 2022-W03 are distributed across 66 countries on all 6 continents, with most resolvers located in \ac{EU} (78, 39.20\%), \ac{NA} (54, 27.14\%), and Asia (35, 17.59\%).
Looking at the distribution over ASes, we find that 55 (27.64\%) are attributed to ANEXIA (AS42473), whose \emph{DoQ resolvers} are all operated by \emph{nextDNS}.
The remaining 144 resolvers (72.36\%) are distributed over 72 ASes, with most ASes hosting 1--2 resolvers.\\

While our measurements show a slowly but steadily increasing adoption of DoQ on resolvers worldwide, the observed week-over-week fluctuations reflect the ongoing development and standardization process with rapidly changing implementations and services.
Considering that these experimental deployments of the resolvers, along with their geographical locations, can substantially affect the overall response times, in particular when multiple RTTs are required, we investigate the measured response times in the following section.

 \section{Response Times}
\label{sec:performance}

To study the response times of DoQ in comparison to \ac{DoUDP}, \ac{DoTCP}, \ac{DoT}, and \ac{DoH}, we issue response time measurements every hour over the course of 2022-W03 (January 17--23), targeting all \emph{DoQ-verified} resolvers identified in 2022-W02 (see \autoref{sec:methodology}).
From these 1,204 resolvers, 1,148 answer our requests via DoQ, 663 via DoH, 1,028 via DoT, 630 via DoTCP, and 455 via DoUDP in 2022-W03.
A total of 264 resolvers, \ie \emph{DoX-verified}, offer all stated DNS protocols simultaneously.
While the adoption measurements (\autoref{sec:adoption}) are independent of the selected vantage point, we acknowledge that the vantage point introduces a location bias for our response time measurements (see \autoref{sec:limitations}).
To counteract these limitations, we restrict our response time analysis to the 264 \emph{DoX-verified} resolvers, enabling a comparative study of all stated DNS protocols.
The geographical distribution of these resolvers follows the distribution observed in the adoption scan:
Asia dominates with 123 (46.59\%) of the \emph{DoX-verified} resolvers, followed by EU with 83 (31.44\%), and NA with 51 (19.32\%).
The remainder are attributed to OC (5 resolvers, 1.89\%) and AF (2 resolvers, 0.76\%).

To account for the different transport protocol mechanisms leveraged by the measured protocols, we differentiate between the \emph{round-trip time (RTT)}, the \emph{resolve time}, and the \emph{handshake time}.
We define the \emph{resolve time} as the time between the moment the first packet of the DNS query is sent until the moment a valid DNS response is received.
Considering we ensure that our requested DNS record is cached by the targeted resolver through cache warming (see \autoref{sec:methodology}), the \emph{resolve time} is, therefore, expected to resemble roughly 1 \emph{RTT} for every measured protocol.
In addition to the \emph{resolve time}, we define the \emph{handshake time} as the time between the moment the first packet of the session establishment is sent until the moment the (encrypted) session to the resolver is established.
Note that since DoUDP uses a connectionless protocol and, therefore, has no connection establishment, it is omitted from the \emph{handshake time} discussion below.
The DoTCP \emph{handshake time} resembles the TCP 3-way handshake, \ie 1 \emph{RTT}.
For DoT and DoH, the TLS handshake is added to the TCP handshake:
Using TLS~1.2, the \emph{handshake time} is 3 \emph{RTTs} for DoT and DoH, which is decreased down to 2 \emph{RTTs} with the usage of TLS~1.3.
Since QUIC combines the handshakes of the connection as well as the encryption in the \texttt{Initial} frame, and we ensure that QUIC's \texttt{Version Negotiation} and \texttt{Address Validation} features do not affect the actual DNS response time measurement query (see~\autoref{sec:methodology}), the \emph{handshake time} of DoQ should resemble 1 \emph{RTT}.

Note that we send every request with a new session for every protocol as a single query; we acknowledge that using a previously established session would reduce the overhead introduced by the \emph{handshake time} for subsequent queries, \eg by using \texttt{edns-tcp-keepalive} on TCP-based sessions.
In addition, the stated \emph{handshake times} can further be optimized \emph{between} sessions by the usage of protocol mechanisms such as \texttt{\acf{TFO}} using TCP, or \texttt{0-RTT} for TLS~1.3 (``early data'') using TCP as well as QUIC (see~\autoref{sec:limitations}).
Hence, we investigate the 1,204 \emph{DoQ-verified} resolvers for the support of \texttt{edns-tcp-keepalive} and \texttt{TFO}, as well as TLS~1.3 \texttt{0-RTT} for QUIC.
We do not explicitly investigate the support of TLS~1.3 \texttt{0-RTT} for TCP, which we instead leave open for future work.
For \texttt{edns-tcp-keepalive}, we find support only on the \emph{AdGuard} resolvers, although they respond with a \texttt{timeout} value of \texttt{0} which instructs the client to directly close the session after having received the response.
As for \texttt{TFO}, we find 208 resolvers supporting the TCP extension, of which no resolver is included in our \emph{DoX-verified} set.
Finally, none of the \emph{DoQ-verified} resolvers offer support for QUIC \texttt{0-RTT}.
However, the lack of 0-RTT support might be a deliberate choice, as the use of 0-RTT exposes clients to privacy risks~\cite{ietf-dprive-dnsoquic}.

\begin{figure}[t]
	\vspace{-1em}
	\centering

	\begin{subfigure}[c]{0.486\linewidth}
		\includegraphics[width=\columnwidth]{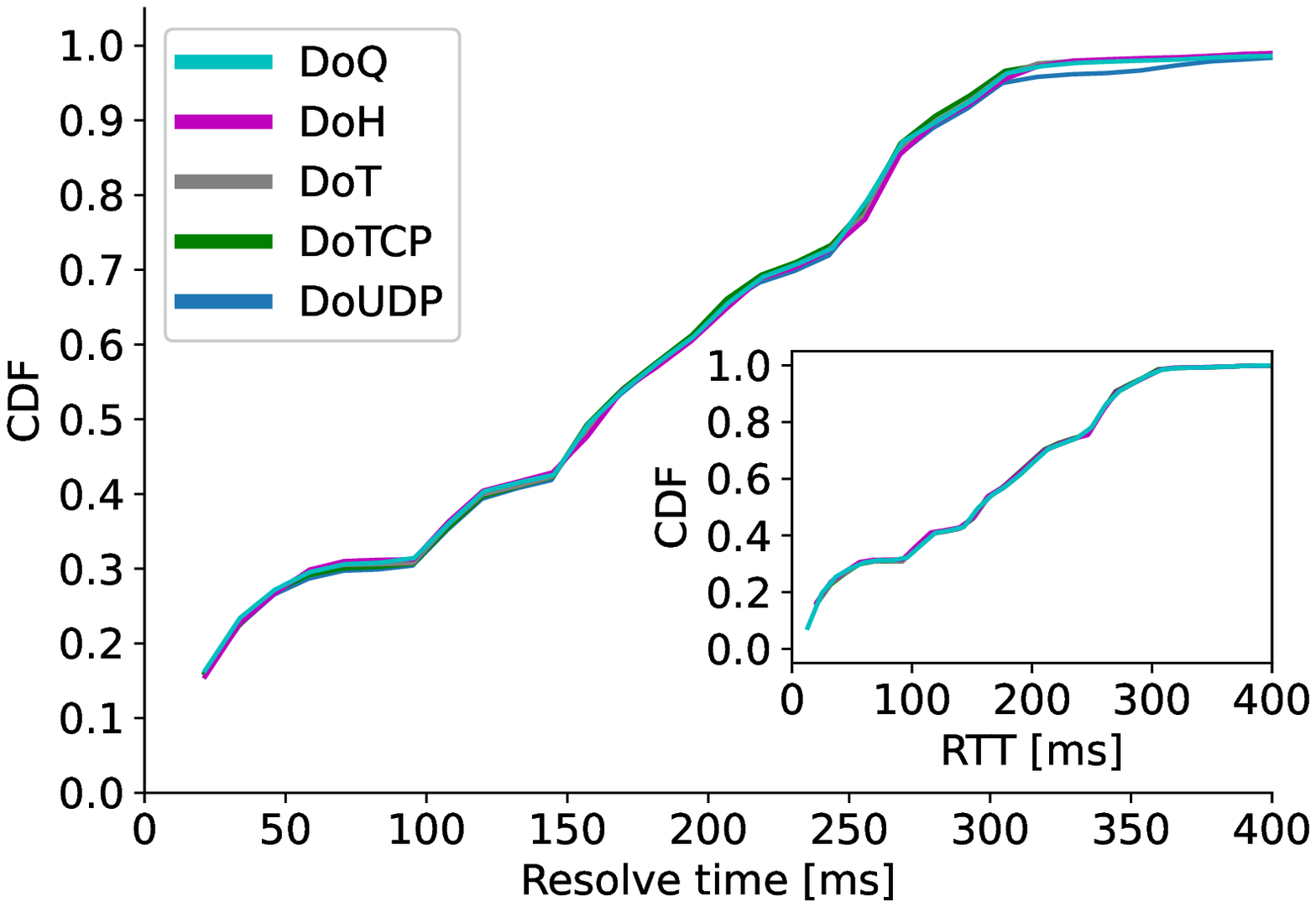}
		\subcaption{\emph{Resolve time} and \emph{RTT} distribution. \phantom{--------------------------}}
	\end{subfigure}
	\hfill
	\begin{subfigure}[c]{0.48\linewidth}
		\includegraphics[width=\columnwidth]{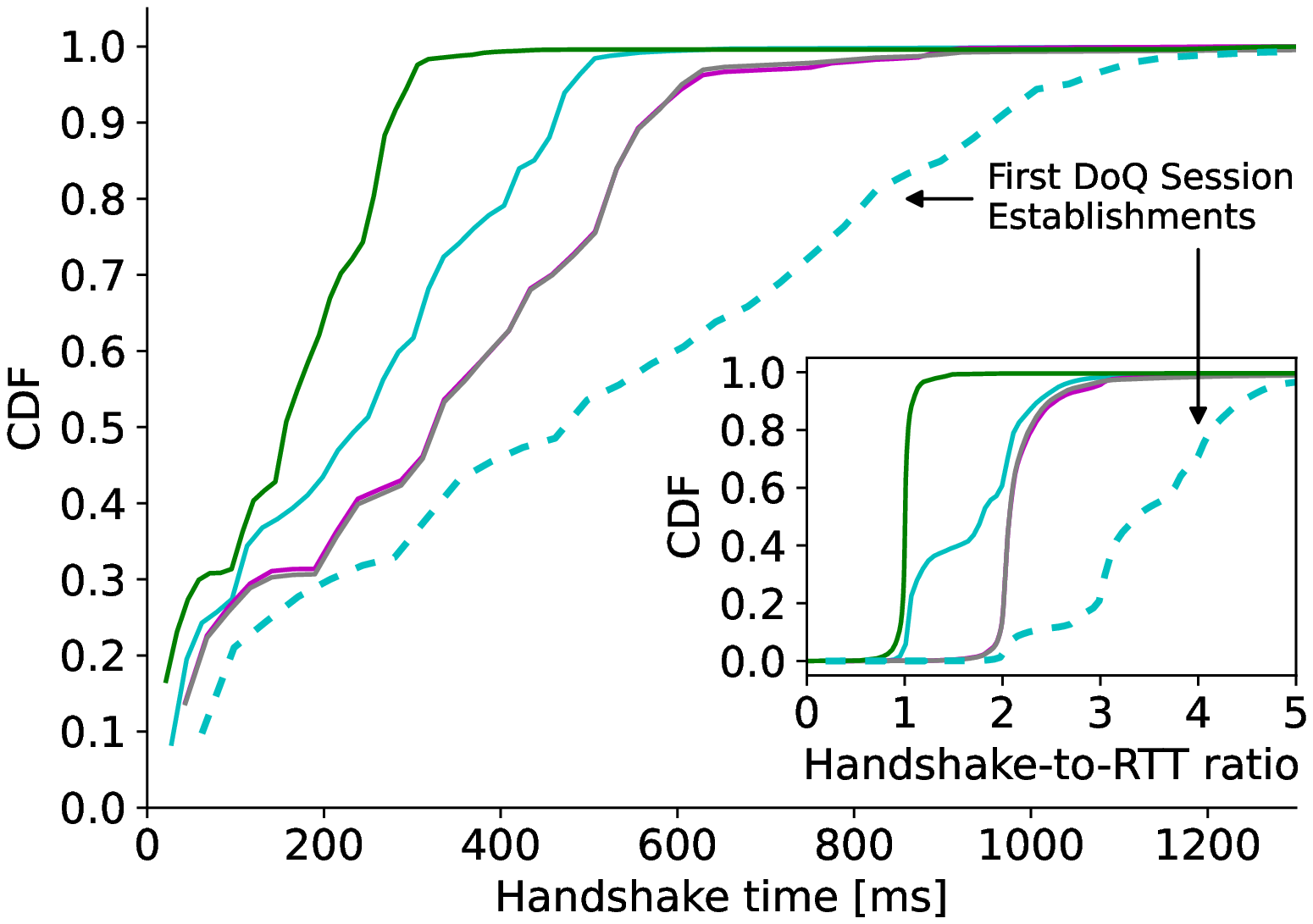}
		\subcaption{\emph{Handshake time} and \emph{handshake-to-\phantom{------}RTT ratio} distribution.}
	\end{subfigure}
	\vspace{-1em}
	\caption{Distributions of response time metrics, targeting 264 \emph{DoX-verified} resolvers. Please note the different $x$-axis scales.
	} \label{fig:multi-cdfs}
	\vspace{-1em}
\end{figure}

Figure \ref{fig:multi-cdfs} presents the distribution of the \emph{resolve time} and the \emph{RTT} (a, left), as well as the \emph{handshake time} and \emph{handshake-to-RTT ratio} (b, right) of the response time measurements toward the 264 targeted \emph{DoX-verified} resolvers; please note the different $x$-axis scales.
The steps in the CDF lines can be explained by two consecutive hops that have a high difference in their latencies, \eg when crossing continental borders.
For our response time analysis, we only consider measurements which successfully return a valid DNS response containing a \texttt{Response Code} within a timeout of 5 seconds.
Further, we limit the analysis of DoQ, DoH, DoT, and DoTCP to measurements for which the corresponding \emph{RTT} measurement is successfully answered by the resolvers.
For DoUDP, we exclude the \emph{RTT} measurements, as they were not replied to by any resolver, but include the \emph{resolve} time for comparison.
Analyzing the \emph{resolve} time in Fig. \ref{fig:multi-cdfs} (a, left), we observe that the distributions of all protocols are almost identical, and match the distributions of the respective \emph{RTT} as shown in the subplot.
Hence, we confirm that the \emph{resolve} time indeed resembles 1 \emph{RTT}, regardless of the protocol.

In contrast to \emph{resolve time} and \emph{RTT}, the \emph{handshake time} presented in Fig.\ref{fig:multi-cdfs} (b, right) shows a vastly different picture.
DoTCP (green line) offers the fastest \emph{handshake} times over all protocols with a median of 156ms (mean 153ms), which is expected due to DoTCP only requiring the TCP handshake (1 \emph{RTT}) for the establishment of the session.
On the other hand, DoT (gray line) and DoH (magenta line) show almost identical \emph{handshake} times, with medians of around 322ms and means of around 315ms.
As both protocols require the TCP handshake plus the TLS handshake for session establishment, the \emph{handshake} times should resemble 3 times the measured \emph{handshake} times of DoTCP when TLS~1.2 is used (3 \emph{RTTs} in total), and 2 times when TLS~1.3 is used (2 \emph{RTTs} in total):
Analyzing the negotiated TLS versions, we observe that 99.6\% of DoT and 96.7\% of DoH measurements use TLS~1.3, whereas the remaining ones use TLS~1.2.
Analyzing DoQ, we find an unexpected result (solid cyan line):
Since QUIC combines the connection and encryption handshakes into 1 \emph{RTT}, DoQ is expected to have the same distribution as DoTCP.
However, with a median of 235ms and a mean of 233ms, the DoQ \emph{handshake} times observed are higher than expected, having its distribution in between the distributions of DoTCP, and DoT and DoH.

To investigate this, we analyze the distribution of the relative number of \emph{RTTs} which are required by the \emph{handshakes} as shown in Fig.\ref{fig:multi-cdfs} (b, right, subplot).
We divide each successful \emph{DoX} \emph{handshake time} measurement by its consecutive \emph{RTT} measurement, thus, showing the distribution of the \emph{handshake-to-RTT ratio} of each measurement pair.
For DoTCP (green line), we observe that the \emph{handshake} resembles 1 \emph{RTT} as expected.
Moreover, DoT (gray line) and DoH (magenta line) again overlap and converge into a long tail, roughly resembling the expected 2 \emph{RTTs} for TLS~1.3 up until the median.
On the other hand, DoQ (solid cyan line) differs drastically from the expected \emph{handshake} of 1 \emph{RTT}:
With around 20\% of measurements showing an \emph{RTT} of 1, DoQ converges to 2 RTTs at the 60\textsuperscript{th} percentile; hence, roughly 40\% of DoQ measurements require more than 2 \emph{RTTs}, which is twice as much as expected in comparison.
To investigate this, we analyze the \texttt{qlog}~\cite{qlog} outputs recorded during our response time measurements, which enable us to analyze the packet exchanges in detail.
Using the \texttt{qlogs}, we confirm that the response time measurements are not affected by QUIC's \texttt{Version Negotiation} feature, as we use the previously negotiated QUIC \texttt{Version} of the cache warming session for the handshake of the subsequent DNS response time measurement session (see~\autoref{sec:methodology}).
However, we attribute the additional 1 \emph{RTT} to the \texttt{Address Validation} feature of QUIC, which is a requirement for every session to prevent traffic amplification attacks by validating that the client is able to receive packets.
To perform \texttt{Address Validation}, the QUIC standard~\cite{rfc9000} defines 1 \emph{implicit} and 2 \emph{explicit} mechanisms, with the \emph{implicit} mechanism validating the address by receiving a packet protected with a handshake key (i.e., 1 additional \emph{RTT}).
The first \emph{explicit} mechanism uses a \texttt{Retry} token sent by a server as a response to the clients \texttt{Initial} frame, instructing the client to re-issue the \texttt{Initial} frame with the server-constructed token (i.e., 1 additional \emph{RTT}).
The second \emph{explicit} mechanism also leverages a server-constructed token: If a server issued a token using a \texttt{New\_Token} frame in a previous session, it can be used in the \texttt{Initial} frame of a subsequent session (i.e., no additional \emph{RTTs}).

Analyzing the \texttt{qlogs}, we find that in every cache warming session a \texttt{Retry} token is sent, and the client is validated using the first \emph{explicit} mechanism.
Moreover, we observe that a \texttt{New\_Token} frame is also issued in every cache warming session, which we use in the subsequent DNS response time measurement session in order to validate the address within the clients first \texttt{Initial} frame.
For those subsequent measurements, we confirm that every DNS response time measurement session is not affected by an additional \texttt{Retry} frame and, thus, no additional \emph{RTT}, as the \texttt{Address Validation} is fulfilled.
However, we find that resolvers still enforce the traffic amplification limit of 3 times the amount of data they received despite successful validation of the client's address:
Depending on the X.509 certificate issued by a server, its size might exceed the traffic amplification limit, which requires the client to \texttt{ACK} data before the server sends the remaining bytes.
Hence, an additional \emph{RTT} is required, resulting in 2 \emph{RTTs} in total as observed in roughly 40\% of DoQ measurements (see Fig.\ref{fig:multi-cdfs}, b, right, cyan lines) -- 2 times as much as expected.

We further analyze the \emph{handshake times} of cache warming queries:
This allows us to investigate the effect of \texttt{Address Validation} mechanisms on the DoQ \emph{handshake time} required for the first session establishment between a client and a resolver (see Fig.\ref{fig:multi-cdfs}, b, right, dashed cyan lines).
With a median of 468ms and a mean of 487ms, the \emph{handshake} times for the first session establishment are roughly doubled in comparison to subsequent sessions.
Analyzing the \texttt{qlogs}, we find that the traffic amplification limit is also enforced in the cache warming sessions following successful \texttt{Address Validation}, which can therefore require up to 4 \emph{RTTs} (i.e., \texttt{Initial}, \texttt{Version Negotiation}, \texttt{Retry}, and \texttt{Amplification Limit}) -- 4 times as much as expected.

Both our DoQ \emph{handshake time} analyses of cache warming and subsequent queries show that an already validated address is still constrained by the traffic amplification limit until the client sends another frame, which adds 1 \emph{RTT} to the handshake.
However, while the QUIC standard states that the traffic amplification limit is to be enforced  \emph{until} a client is successfully validated, we argue that our observations are most likely an unintentional effect of the QUIC implementations used by the DoQ resolvers.
Hence, we suggest resolvers to not enforce the traffic amplification limit on already validated client addresses to optimize the performance, which results in a reduction by 1 \emph{RTT} during the \emph{handshake}.

 \section{Limitations and Future Work}
\label{sec:limitations}

We acknowledge that the selected vantage point introduces a location bias for our measurements,
in particular for the measured latencies in the response time analysis (\autoref{sec:performance}). The highly varying geographical distances to the resolvers (whose distribution exhibits further biases, see \autoref{sec:adoption}) inherently affect the delays, especially if multiple round-trips are required.
Hence, we plan to address this limitation by performing measurements from distributed vantage points worldwide to obtain a more representative view on \ac{DoQ} response times around the globe.

Moreover, we acknowledge that public DNS resolvers often leverage IP anycast, about which we could not find any publicly available information for DoQ resolvers of \emph{AdGuard} and \emph{nextDNS}.
In addition, by cross-referencing anycast IP addresses of public DNS providers used in related work measuring DoT and DoH~\cite{lu2019dns,doan2021measuring,garcia2021large}, we were not able to identify these public DNS providers within our set of \emph{DoQ-verified} resolvers.

Further, we miss resolvers that do not accept \ac{DoQ} requests without \ac{SNI} information.
For instance, Google requires queries over TLS~1.3 (which, thus, also affects QUIC) to use the SNI extension~\cite{googlepublicdns}.
As a result, \ac{DoQ} queries to 8.8.8.8 are not responded to by Google, whereas queries with the \texttt{HostName} set to \texttt{dns.google.com} in the \ac{SNI} extension do trigger a DNS response.
Since we do not include \ac{SNI} in our requests due to not knowing the corresponding \texttt{HostName} for every identified resolver, we cannot identify and measure resolvers with such requirements.
Therefore, the list of \ac{DoQ} resolvers measured in our study is not exhaustive, as we only consider open resolvers that do not require \ac{SNI}.
Moreover, we only consider IPv4 resolvers in our study; future work should also consider scanning the IPv6 address space as a complement, e.g., based on IPv6 hitlists~\cite{GasserSFLKSHC18}.

Finally, we plan to further evaluate DoQ by using previously established sessions for subsequent queries, as well as TLS~1.3 \texttt{0-RTT} between sessions in a future study: Both mechanisms reduce the overhead introduced by the \emph{handshake time}, which affects application layer protocols that typically require multiple DNS queries in rapid succession.
 \section{Conclusion}
\label{sec:conclusion}

DNS over QUIC promises to improve on the established encrypted DNS protocols by leveraging the QUIC transport protocol.
In our study, we detailed a slowly but steadily increasing adoption of DoQ on resolvers worldwide, where the observed week-over-week fluctuations reflect the ongoing development and standardization process with rapidly changing implementations and services.
Analyzing the response times of DoQ, we showed that the DoQ \emph{handshake} times fully utilize QUIC's potential in around 20\% of measurements.
However, roughly 40\% of measurements show considerably higher \emph{handshake} times than expected, which traces back to the enforcement of the traffic amplification limit despite successful validation of the client's address.
While this shows still unused optimization potential, DoQ already outperforms DoT as well as DoH, making it the best choice for encrypted DNS to date.

In conclusion, our study provided a first look at DNS over QUIC.
However, we presented only a glimpse of the potential of DoQ: With the expectation that the upcoming standardization of DoQ will cause a surge in adoption along with optimizations of existing implementations, future studies will reveal whether DoQ will truly become the \emph{``One to Rule them All''}. 
\section*{Acknowledgements}
We thank Luca Schumann and Simon Zelenski for their valuable support, as well as Jan Rüth and the anonymous reviewers for their insightful feedback.

\bibliographystyle{splncs04}
\bibliography{index}

\end{document}